\newcommand\lefp{\left(}
\newcommand\rigp{\right)}
\newcommand\lefq{\left[}
\newcommand\rigq{\right]}

\newcommand\sch{Schwarzschild}

\newcommand\beque{\begin{equation}}
\newcommand\beq{\begin{equation}}
\newcommand\eeq{\end{equation}}
\newcommand\eeque{\end{equation}}

\newcommand\beqnl{\begin{eqnarray}}
\newcommand\beqna{\begin{eqnarray*}}

\newcommand\eeqna{\end{eqnarray*}}
\newcommand\eeqnl{\end{eqnarray}}

 \def\NN{\hbox{\sf I\kern-.13em\hbox{N}}}
 \def\HH{\hbox{\sf I\kern-.13em\hbox{H}}}
 \def\DD{\hbox{\sf I\kern-.13em\hbox{D}}}
 \def\RR{\hbox{\sf I\kern-.14em\hbox{R}}}
 \def\CC{\hbox{\sf I\kern-.44em\hbox{C}}}
 \def\ZZ{{\hbox{\sf Z\kern-.43emZ}}}
 \def\QQ{\hbox{\sf C\kern -.48emQ}}
 \def\Cc{\hbox{\sf C\kern -.47em {\raise .48ex \hbox{$\scriptscriptstyle |$}}
   \kern-.5em {\raise .48ex \hbox{$\scriptscriptstyle |$}} }}
 \def\Qq{\hbox{\sf Q\kern -.57em {\raise .48ex \hbox{$\scriptscriptstyle |$}}
   \kern-.55em {\raise .48ex \hbox{$\scriptscriptstyle |$}} }}

\documentstyle[aps,prd,preprint]{revtex}
\tightenlines
\begin{document}
\draft
%@twocolumnfalse\endcsname

%\rightline{gr-qc/9811060}
\vskip1pc

\title{ Quantum Properties of the Electron Field in 
Kerr--Newman Black Hole Manifolds}
\author{F. Belgiorno\footnote{E-mail: belgiorno@mi.infn.it}}
\address{Dipartimento di Fisica, Universit\`a di Milano, 20133 
Milano, Italy, and\\
I.N.F.N., Sezione di Milano, Italy}
\author{M. Martellini\footnote{E-mail: maurizio.martellini@mi.infn.it}}
\address{Dipartimento di Fisica, Universit\`a di Milano, 20133 
Milano, Italy\\ 
I.N.F.N., Sezione di Milano, Italy, and\\
Landau Network at ``Centro Volta'', Como, Italy}

\date{\today}
\maketitle
\vskip -0.7 truecm
\begin{abstract}
 
\vskip -0.3 truecm
We study some spectral features of the one--particle electron Hamiltonian 
obtained by separating the Dirac equation 
in a Kerr--Newman black hole background. We find that the essential 
spectrum includes the whole real line. As a consequence, 
there is no gap in the spectrum and discrete eigenvalues are 
not allowed for any value of the  black hole 
charge $Q$ and angular momentum $J$. 
Our spectral analysis will be also related to 
the dissipation of the black hole angular momentum and charge. 

\vskip -0.2 truecm
\end{abstract}

\vskip0.2cm\noindent
\\PACS:  04.62.+v, 02.30.Tb, 04.70.Dy \\
Keywords: Dirac equation, Quantum field theory in Curved Spacetime

\vskip2pc
%]

\newpage

\section{Introduction}

In this letter we show that in a Kerr--Newman black hole background 
the reduced one--particle Hamiltonian for the electron field, 
obtained from the Dirac equation by separation of variables 
\cite{page,cartereq,pekeris,jmp}, is characterized by  
a peculiar spectral feature: Its essential spectrum is given by $\RR$.  
As a consequence, the absence of gap in the spectrum and of 
discrete eigenvalues is deduced. 
The above results hold for any value of $Q,J$. 
We will study the 
general Kerr--Newman case. The Reissner--Nordstr\"{o}m 
case $J=0,Q\not =0$ has been studied in \cite{belgio} 
and here we extend the results of 
\cite{belgio} to the Schwarzschild case 
$J=0,Q=0$\footnote{Some of our results for Reissner--Nordstr\"{o}m 
represent a rigorous implementation of the ones found 
in \cite{soffart}, which are based on a WKB approximation.}.\\ 
Some qualitative considerations about the black hole 
loss of charge and/or angular momentum are also allowed by our 
spectral analysis. 
As it is well known, the presence of the 
Klein paradox \cite{thaller,soffbook} 
has been related to the quantum dissipation 
of the black hole charge and angular momentum 
\cite{gibbons,damo,deruelle,ruffini}. The conditions for the 
existence of the Klein paradox are studied. A further discussion is found in 
the conclusions.

\section{One--particle electron Hamiltonian in Kerr--Newman 
black hole manifolds}

The metric of a Kerr--Newman black hole of mass $M$, angular 
momentum $J$ and charge $Q$ in Boyer--Lindquist coordinates 
$(t,r,\theta,\phi)$ is
\cite{wald,carter} 
\beqna
ds^2&=&- \lefp \frac{\Delta-a^2 \sin^2 \theta}{\Sigma} \rigp dt^2
-\frac{2 a \sin^2 \theta \lefp r^2+a^2-\Delta \rigp}{\Sigma} dt 
d\phi\\
&+&\lefq \frac{(r^2+a^2)^2-\Delta a^2 \sin^2 \theta}{\Sigma} 
\rigq \sin^2 \theta d\phi^2 +\frac{\Sigma}{\Delta}dr^2 +\Sigma 
d\theta^2 
\eeqna
where $a=J/M$, $\Sigma\equiv r^2+a^2 \cos^2 \theta$ and $\Delta\equiv 
r^2+a^2+Q^2-2Mr=(r-r_{+})(r-r_{-})$; 
$r_{+} \geq r_{-}>0$ correspond to the event horizon radius and to 
the Cauchy horizon radius respectively. We will consider only the external 
region $r\in (r_{+},\infty)$. The electromagnetic 
vector potential is $
A_{\mu}=(Q r)/\Sigma (-1,0,0, a\; \sin^2 \theta)$. 
The radial part of the separated Dirac equation for a charged field 
\cite{page,cartereq,pekeris,jmp} 
in the case of the electron field (mass $m_e$, charge $-e$) is given by 
\beqna
(D_{0}- i \frac{-e Q r}{\Delta})R_{-}(r)&=&(\lambda+i m_{e} r) R_{+}(r)\cr
(\Delta D^{\dagger}_{\frac{1}{2}}+ i (-e Q r))R_{+}(r)&=&(\lambda-i m_{e} r) 
R_{-}(r)
\eeqna
where $D_{0}=\partial_r+ i K/\Delta$; 
$D^{\dagger}_{\frac{1}{2}}=\partial_r - i K/\Delta+(r-M)/\Delta$; 
$K=(r^2+a^2) \omega+a m$. 
$\omega$ appearing in $K$ represents the one--particle energy of the solution 
of the separated Dirac equation, $\lambda$ is the eigenvalue for the 
$\theta$-dependent part of the spinor and $m$ is the semi--integer 
azimuthal quantum number. 
A rather straightforward algebraic manipulation of the above equations, 
obtained by posing $R_{-}=F+i G$, $R_{+}= (F-i G)/\sqrt{\Delta}$ (cf. 
\cite{pekeris}), allows to get the following system:
\beqna
\frac{dF}{dr}&=&+\frac{\lambda}{\sqrt{\Delta}} F + (
\frac{K}{\Delta}+\frac{m_e r}{\sqrt{\Delta}} +
\frac{-e Q r}{\Delta} )G\cr
\frac{dG}{dr}&=&-\frac{\lambda}{\sqrt{\Delta}} G + (
-\frac{K}{\Delta}+\frac{m_e r}{\sqrt{\Delta}} -
\frac{-e Q r}{\Delta} )F.
\eeqna 
We define, by means of the change of variable 
$dx/dr=\delta/\Delta$, where $\delta\equiv r^2+a^2$, 
the tortoise--like coordinate $x$; in the non extremal case one gets 
$x=r+(r_{+}^2+a^2)/(r_{+}-r_{-})\cdot \log((r-r_{+})/r_{+})      
-(r_{-}^2+a^2)/(r_{+}-r_{-})\cdot \log((r-r_{-})/r_{-})$
and in the extremal one   
$x=r+2 r_{+}\cdot \log((r-r_{+})/r_{+})-(r_{+}^2+a^2)/(r-r_{+})$. 
In both cases, $x\in (-\infty,+\infty)$. 
Then, it is easy to show that the above system is equivalent to the 
eigenvalue problem 
\beque
H \vec{W}=\omega \vec{W}
\label{eigen}
\eeque
where $\vec{W}\equiv (F,G)^T$; $H$ is 
the one--particle (reduced) Hamiltonian in matrix form
\beque
H=H_{0}+V(r(x))
\label{hamilt}
\eeque
with 
\[ H_{0}=\left[
\begin{array}{cc}
0 &  - \partial_x \cr 
\partial_x &\ 0\end{array} \right] 
\]
and 
\[ V(r(x))=\left[
\begin{array}{cc}
\frac{-a m+m_e r \sqrt{\Delta} - e Q r}{\delta} &  
-\lambda \frac{\sqrt{\Delta}}{\delta} \cr 
-\lambda \frac{\sqrt{\Delta}}{\delta} 
&\frac{-a m - m_e r \sqrt{\Delta} - e Q r}{\delta} \end{array} \right]. 
\]
The Hamiltonian (\ref{hamilt}), which is the one--particle energy operator 
calculated by separation of variables, in the following will be also 
called partial wave operator. We define 
$\Omega\equiv a/\delta;\ \Phi\equiv (Q r/\delta)$; these 
quantities, valued at the event horizon, become 
the black hole angular velocity $\Omega_h$ and the black hole 
electric potential $\Phi_h$ respectively. Then the potential 
can be rewritten as 
\[ V(r(x))=\left[
\begin{array}{cc}
-m \Omega-e \Phi+\frac{m_e r \sqrt{\Delta}}{\delta} &  
-\lambda \frac{\sqrt{\Delta}}{\delta} \cr 
-\lambda \frac{\sqrt{\Delta}}{\delta} 
&-m \Omega-e \Phi-\frac{m_e r \sqrt{\Delta}}{\delta} \end{array} \right]. 
\]
The properties of the operator $H$ can be analyzed in the 
framework of the so called Dirac systems of ordinary differential 
equations \cite{weidmann}. It is easy to show that $H$ is 
formally self-adjoint in the Hilbert space $L^2 (\RR,dx)^2$ and 
that it is also essentially self-adjoint in 
$C_0^{\infty}(\RR)^2$. Indeed, the so called limit point case 
holds at $-\infty$ and at $+\infty$ (cf. corollary to the theorem 
6.8 in \cite{weidmann}) and so its deficiency indices are $(0,0)$ 
(cf. theorem 5.7 of \cite{weidmann}). This means that there is no 
need to select boundary conditions for the electron 
one--particle Hamiltonian on the given manifold.   

\section{essential spectrum}

We show that the discrete spectrum of 
the reduced one--particle Hamiltonian is empty and that there is no gap in the 
spectrum, as a consequence of the fact 
that the essential spectrum $\sigma_e$ coincides with 
$\RR$. The decomposition method \cite{weidmann} and theorems 
16.5 and 16.6 of \cite{weidmann}\footnote{Cf. also 
\cite{weidart}, the theorems above correspond to Korollar 6.9 and 
Satz 6.10 respectively.} allow 
us to find out the essential spectrum of the reduced Hamiltonian $H$. 
An analogous study was carried 
out in the Reissner--Nordstr\"{o}m case in \cite{belgio}, 
where the absence of gap and of the discrete spectrum 
was implicitly shown for Reissner-Nordstr\"{o}m 
black holes. Here the same reasoning is applied to the general 
Kerr--Newman case. Physical consequences of our spectral analysis 
will be also considered and a second quantization formalism will be 
understood in the following section. Moreover, 
it is necessary to recall that the correct physical 
state for quantum field theory in a non extremal black hole background 
has to be selected according to suitable analyticity requirements 
for the metric and the fields on the extended manifold \cite{HH}. 
For the (eternal) Kerr--Newman non extremal case 
the one--particle Hamiltonian vacuum is not the correct physical  
Hartle--Hawking state; nevertheless, it is still 
related to the physical state: On the extended 
\sch\ background its ``heating up'' at the black 
hole temperature in the scalar field case gives rise 
to the Hartle--Hawking state \cite{kay}. Cf. also \cite{israel,frolov} for 
the scalar field case in the Kerr background. 
Extremal black holes cannot a priori be treated on the same 
foot as the non extremal ones, because of the lack of a geometric 
temperature. Cf. also \cite{vanzo}. 
The correct physical state is not known. We choose 
conservatively the Boulware (one--particle Hamiltonian) vacuum.\\ 
Before applying the mathematical tools cited above, 
it is useful to introduce the following constant shift in 
$\omega$: $\omega\to \omega -e \Phi_{h}-m \Omega_h$ in such a way that the 
eigenvalue problem (\ref{eigen}) becomes an eigenvalue problem (with 
eigenvalue $\omega$) for the Hamiltonian (\ref{hamilt}) 
with the shifts 
$e \Phi\to e (\Phi-\Phi_{h});\ m \Omega \to m (\Omega-\Omega_{h})$ 
in the potential. We will call the shifted potential again $V$. 
Our choice is purely conventional. See also the discussion in the 
following section.\\ 
According to the decomposition method, in order to locate 
the essential spectrum $\sigma_e$ of $H$, it is sufficient to locate the 
essential spectrum of the restrictions of $H$ to suitable 
subintervals of $\RR$. Particularly, let $H_{-}$ and $H_{+}$ be 
self--adjoint extensions of the reduced 
Hamiltonian restricted to the intervals $(-\infty,0]$ and $[0,+\infty)$ 
respectively. Then it can be proved that 
$\sigma_{e}(H)=\sigma_{e}(H_{-})\cup \sigma_{e}(H_{+})$ 
(cf. theorem 11.5 of \cite{weidmann}). 
In order to apply theorem 16.5 of \cite{weidmann},  
we determine the limits of the potential $V$ for $x\to +\infty$ 
and $x\to -\infty$: 
\[ \lim_{x\to +\infty} V(r(x))\equiv V_{+}=\left[
\begin{array}{cc}
m_e +m \Omega_h+ e \Phi_h& 0\cr 
0 & -m_e +m \Omega_h+e \Phi_h\end{array} \right] 
\]
and
\[ \lim_{x\to -\infty} V(r(x))\equiv V_{-}=\left[
\begin{array}{cc}
0 & 0\cr 
0 & 0 \end{array} \right]. 
\]
Indeed, theorem 16.5 of \cite{weidmann} states that, 
if the $w_{-}\leq w_{+}$ are the 
eigenvalues of $V_{+}$ (analogously for $V_{-}$), 
there is no essential spectrum contribution 
from the open interval $(w_{-},w_{+})$. Then in the case of $H_{+}$ 
one gets $\sigma_{e}(H_{+})\cap (-m_e+e \Phi_{h}+m \Omega_h,m_e+e 
\Phi_{h}+m \Omega_h)=\emptyset$. 
(For $H_{-}$ the following trivial result holds:  
$\sigma_{e}(H_{-})\cap \emptyset=\emptyset$). 
The above result holds 
in the non extremal case as well as in the extremal one. 
Moreover, one can apply theorem 16.6 of \cite{weidmann}  
that will enable us to locate exactly the essential spectrum 
contribution arising from near $+\infty$ and from 
near the horizon. We first study $H_{+}$ near $x=+\infty$. 
According to theorem 16.6 of \cite{weidmann}, if 
$|\cdot|$ stays for a norm in $\CC^{2 \times 2}$ and 
if for some $d\in (0,+\infty)$ the limit 
$\lim_{x\to +\infty} \frac{1}{x} \int_{d}^{x} ds |V(r(s))-V_{+}|$ 
is zero then the essential spectrum of $H_{+}$ 
contains the complement of the open interval $(w_{-},w_{+})$ 
(of course, the above limit is trivially zero if the  
integral is finite for $x\to +\infty$). It is easy to show that 
the above integrand near $+\infty$ behaves as $\frac{1}{r}$ 
and so $\lim_{x\to +\infty} \frac{1}{x} \int_{d}^{x} ds 
|V(r(s))-V_{+}|=0$, as can be also verified by using L'Hospital's rule. 
Then one can conclude that 
$\sigma_{e}(H_{+})\supset (-\infty,-m_e+e \Phi_{h}+m \Omega_h] 
\cup [m_e+e \Phi_{h}+m \Omega_h, +\infty)$.
In the case of $H_{-}$, the study is analogous but it is 
necessary to distinguish the non extremal case and the extremal 
one. By defining the variable $y=-x$ one can apply theorem 
16.6 of \cite{weidmann} by evaluating  
$\lim_{y\to +\infty} \frac{1}{y} \int_{\alpha}^{y} dy |V(r(y))-V_{-}|=
\lim_{r\to r_{+}} \frac{1}{y(r)}\int_{r}^{\beta}  dr 
\frac{\delta}{\Delta} |V(r)-V_{-}|$. 
The integrand is singular as $r\to r_{+}$. 
In the non extremal case one gets that the integral behaves 
as $(r-r_{+})^{-\frac{1}{2}}$ and so there is an integrable 
singularity. Then the above limit is trivially $0$. 
In the extremal case the integrand behaves as  
$(r-r_{+})^{-1}$ and the most singular contribution of 
the integral behaves as $\log(r-r_{+})$. 
The factor $\frac{1}{y}$ behaves as $r-r_{+}$ in the limit 
$r\to r_{+}$ and so 
one gets  $\lim_{r\to r_{+}} \frac{1}{y(r)}\int_{r}^{\beta}  dr 
\frac{\delta}{\Delta} |V(r)-V_{-}|=0$. Then in both cases 
$\sigma_{e}(H_{-})\supset \RR$. 
All the above results allow us to conclude that  
\beqna
\sigma_{e}(H_{+})&=&(-\infty,-m_e+e \Phi_{h}+m \Omega_h]\cup[m_e+e 
\Phi_{h}+m \Omega_h,+\infty)\cr
\sigma_{e}(H_{-})&=&\RR. 
\eeqna
Then it is evident that $\sigma_{e}(H)=\RR=\sigma (H)$ 
and so one has to conclude that the discrete spectrum is void and that 
there is no gap in the spectrum even if the Dirac field is massive.

\section{Qualitative physical consequences}

We recall that the overlap between negative (positive) asymptotic states 
at $+\infty$ and positive (negative) asymptotic states at 
$-\infty$ has been related with the charge and/or angular 
momentum loss of the black hole 
\cite{gibbons,damo,deruelle,ruffini}. Such an overlap gives 
rise to the Klein paradox and will be called ``Klein 
region"; the related pair creation phenomenon \cite{damo} 
will be called ``Klein effect''; 
``Klein condition'' will be the condition ensuring the Klein paradox
\footnote{The Klein condition 
in the scalar field case identifies the so called superradiant modes. 
But we recall that the superradiance phenomenon 
is not possible for Dirac fields.}; 
cf. also \cite{belgio} and references 
therein for a more detailed discussion. Our analysis represents a 
rigorous implementation for the case of the electron field of the 
analysis given in \cite{gibbons,damo,deruelle,ruffini}. 
Note also that our shift in the potential does not affect the Klein condition
\footnote{Our shift implies 
that the positive states $\omega>m_e+e \Phi_{h}+m \Omega_h$ and 
the negative states  
$\omega<-m_e+e \Phi_{h}+m \Omega_h$ at $+\infty$ have to 
be compared with the positive states $\omega>0$ and the negative 
states $\omega<0$ at the horizon. If the shift is not implemented, 
then the positive states $\omega>m_e$ and the 
negative states $\omega<-m_e$ at $+\infty$ have to 
be compared with the positive states $\omega>-e \Phi_{h}- m \Omega_{h}$ 
and the negative states $\omega<-e \Phi_{h}- m \Omega_{h}$ at 
the horizon. In other words, the extremes of the gap in the spectrum 
at $+\infty$ ``coalesce'' into the single point $0$ at the horizon 
in the shifted potential case and into the point 
$-e \Phi_{h}- m \Omega_{h}$ at the horizon in the non shifted one. 
The Klein condition is the same in both cases.}. We will limit 
ourselves to qualitative considerations. For an explicit calculation of 
the effective action see e.g.
\cite{gibbons,damo,deruelle,ruffini,dewitt,weems}.\\ 
We will start our discussion by summarizing 
the Reissner--Nordstr\"{o}m case; then a study of the 
Kerr case will precede the analysis of the Kerr--Newman one. 

\subsubsection{J=0 cases}

The case $J=0, Q\not = 0$ has been analyzed in \cite{belgio}. 
Therein it is shown that for all the partial wave operators 
the same Klein condition $m_e<e \Phi_h$ (for $Q>0$) holds. 
Then a positive charge flow towards $+\infty$, signalling a 
discharge of the black hole, is possible. 
We stress that the Schwarzschild case $J=0, 
Q=0$ is a particular case of the above result. There is no 
possibility to find a Klein region but also in the 
Schwarzschild case one gets $\sigma_e=\RR=\sigma$. The disappearance 
of the mass gap in the massive Klein--Gordon case in a \sch\ background 
is discussed in \cite{kaycmp}. Cf. also \cite{kayproc}.

\subsubsection{Kerr case}

For the sake of definiteness, we will assume $\Omega_h >0$. 
We deduce that, 
if $|m| \Omega_{h}<m_e$, there is no overlap of the asymptotic 
negative energy states at $+\infty$ and the positive energy 
states at $-\infty$; 
if $|m| \Omega_{h}>m_e$, then a Klein region exists: 
for $m>0$ one gets an overlap of the asymptotic negative energy states at 
$+\infty$ and the positive energy states at $-\infty$; if $m<0$ 
then asymptotic positive states at $+\infty$ overlap asymptotic 
negative states at $-\infty$. It follows that the loss 
of angular momentum of an uncharged Kerr black hole 
in line of principle can take place also by mean of 
charged currents. Indeed, when $m \Omega_h<-m_e$, there is 
a negative charge flow $j_{-}$ towards $+\infty$.
In the case $m \Omega_h >m_e$, there is 
a positive charge flow $j_{+}$ towards $+\infty$.  Globally 
$j_{+}+j_{-}=0$ so no net black hole charge can arise. 
Qualitatively one expects that to a lower value of $|m|$ corresponds a 
bigger contribution to the effective action for the 
angular momentum loss of the black hole (cf. e.g. \cite{dewitt} for the 
massless scalar field case). For an estimate of $|m|$ see the 
appendix.\\
We note that there are remarkable differences with respect to 
the Reissner--Nordstr\"{o}m case. Indeed, in the Kerr case the 
Klein paradox exists 
whichever physical parameters one has, because the Klein condition 
depends on $m$ and definitively there are partial wave 
operators implementing it and all but for a finite number of 
partial wave operators (that are symmetrically placed in $m$ with respect 
to $0$) are affected by a Klein overlap region. 
Moreover the Klein region size 
increases with $|m|$, whereas in the Reissner--Nordstr\"{o}m 
case it is the same for all 
the partial wave operators.

\subsubsection{Kerr--Newman case}

Here we choose $Q>0,\ \Omega_h>0$. A Klein region exists 
if $m<-(m_e+e \Phi_h)/\Omega_h\equiv n_{-}$; 
$m>+(m_e-e \Phi_h)/\Omega_h\equiv n_{+}$. 
The former condition means to get $m<n_{-}<0$ and 
a negative charge flow $j_{-}$ towards $+\infty$; 
the latter implies 
$n_{+}>0$ if $m_e>e \Phi_h$ and $n_{+}<0$ if $m_e<e \Phi_h$ and 
a positive charge flow $j_{+}$ towards infinity. We note that 
$|n_{-}| >|n_{+}|$, so that more partial waves 
contribute to the positive charge flow leaving the black hole:  
The Klein condition is asymmetric in such a way as to couple the loss of 
angular momentum with a loss of charge 
and a net current leaving the black hole takes place. The 
asymmetry is stronger when $m_e<e \Phi_h$ (note that this is the 
discharge condition for a Reissner--Nordstr\"{o}m black hole) 
because in this case 
$j_{+}$ gets contributions from the lowest values of $|m|$. 
For an explicit evaluation of $n_{+},n_{-}$ see the appendix. 
Summarizing, the introduction of a positive charge of the black 
hole originates a net charge flux leaving the black hole, 
associated with a contextual loss of angular momentum: $j_{+}+j_{-}>0$.

\section{conclusions}

The one--particle Hamiltonian for charged Dirac particles 
around a black hole of the Kerr--Newman 
family has been shown to be characterized by the absence of 
discrete eigenvalues and of gap in its spectrum. 
The peculiar properties of the essential spectrum 
contribution from near the event horizon represent the 
reason for such a behavior. 
As a consequence, we can note that, even 
in the case the Hawking 
effect and/or the Klein effect are strongly suppressed, 
no stationary quantum mechanical orbits 
around a black hole can exist. In particular, a 
charged black hole cannot 
form a sort of ``exotic'' atomic system having the  
charged black hole as its nucleus and around an electronic 
cloud, at least as far as the external field approximation 
holds. Such a possibility 
is instead left open e.g. in the case of 
a naked Reissner--Nordstr\"{o}m singularity \cite{belmart}.\\
The conditions allowing the presence of the Klein paradox, which 
is related with the angular momentum and charge dissipation, 
have been also determined. 

\section*{appendix}
 
It is interesting, for an evaluation of $n_{+},n_{-}$,   
to restore the physical dimensions: 
\beqna
n_{+}&=&+\frac{m_e c^2}{\hbar c} \left( \frac{c}{\Omega_h 
r_{+}}\right) r_{+} 
(1-\gamma \Phi^{\ast}_{h})\cr
n_{-}&=&-\frac{m_e c^2}{\hbar c} \left( \frac{c}{\Omega_h 
r_{+}}\right)  r_{+} 
(1+\gamma \Phi^{\ast}_{h})
\eeqna
where we have defined the adimensional factors 
$\gamma\equiv e/(m_{e} \sqrt{G})$ and $\Phi^{\ast}_{h}\equiv 
(\sqrt{G} \Phi_{h})/c^2$. In the case of the electron field 
one gets $\gamma \sim 10^{21}$ and $(m_e c^2)/(\hbar c)\simeq 
2.59\cdot 10^{12}$(meters)$^{-1}$. The following 
inequalities can be useful: $\Phi^{\ast}_{h}<1$; 
$\Omega_{h} r_{+}<c/2$. 
Then $c/(\Omega_h r_{+})\in (2,+\infty)$. Moreover, the introduction of  
the Planck length $l_{pl} =1.6\cdot  10^{-35}$(meters) allows us to write    
$n_{+}=+ 4.18\cdot  10^{-23}\ \left(c/(\Omega_h r_{+})\right)\cdot 
(r_{+}/l_{pl})\cdot (1-\gamma \Phi^{\ast}_{h})$; 
$n_{-}=- 4.18\cdot  10^{-23}\ 
\left(c/(\Omega_h r_{+})\right)\cdot (r_{+}/l_{pl})\cdot 
(1+\gamma \Phi^{\ast}_{h})$. 
In the case of a Kerr black hole ($Q=0$) one 
gets $n_{+}\geq+ 8.36\cdot  10^{-23}\ r_{+}/l_{pl}$; 
$n_{-}\leq- 8.36\cdot  10^{-23}\ r_{+}/l_{pl}$ 
and it results that, even in the case of a very fast rotation, 
only for microscopic black holes with $r_{+}\leq r_{0}\sim 10^{-12}$ 
meters a low $|m|$ contribution to the loss of 
angular momentum by means of the massive electron field is possible. 
In the astrophysical case of an event horizon radius order of 
the Sun Schwarzschild radius one finds $|m|\geq |m_0|\sim 10^{16}$. 
For a slow rotation $|m|$ 
becomes obviously bigger. As far as the general Kerr--Newman case is 
concerned, we note that for $1\gg \gamma \Phi^{\ast}_{h}$ the behavior 
is nearly the same as for a Kerr black hole so that an 
astrophysical black hole will involve in the dissipation of its 
charge and angular momentum only big values of $|m|$. If instead 
$1\ll \gamma \Phi^{\ast}_{h}$, then all the small values of 
$|m|$ are involved in the discharge process however slow the 
rotation of the hole may be.

\end{document}